\documentclass[twocolumn,superscriptaddress,pre]{revtex4-2}

\usepackage{mathpazo}
\usepackage[T1]{fontenc}
\usepackage[latin9]{inputenc}
\usepackage{color}
\usepackage{amsmath}
\usepackage{amssymb}
\usepackage{graphicx}
\usepackage[unicode=true,
 bookmarks=false,
 breaklinks=true,pdfborder={0 0 0},pdfborderstyle={},backref=false,colorlinks=true]
 {hyperref}
\hypersetup{citecolor=blue,filecolor=black,linkcolor=red,urlcolor=blue}

\makeatletter

\newcommand{\be}{\begin{equation}}
\newcommand{\ee}{\end{equation}}

\newcommand{\PC}[1]{\ensuremath{\left(#1\right)}}

\usepackage{breakurl}

\makeatother

\begin{document}
\title{Analyzing lump-type solutions in scalar field models \\
 through configurational information measure}
 
\author{Marcelo A. Feitosa}
\address{Instituto de Física, Universidade Federal de Goiás, 74.690-900, Goiânia,
Goiás, Brazil}

\author{Wesley B. Cardoso}
\email{wesleybcardoso@ufg.br}
\address{Instituto de Física, Universidade Federal de Goiás, 74.690-900, Goiânia,
Goiás, Brazil}

\author{Dionisio Bazeia}
\address{Departamento de Física, Universidade Federal da Paraíba, João Pessoa,
Paraíba, Brazil}

\author{Ardiley T. Avelar}
\address{Instituto de Física, Universidade Federal de Goiás, 74.690-900, Goiânia,
Goiás, Brazil}

\begin{abstract}
In this paper we employ a configurational information measure, specifically
the differential configurational complexity (DCC), to quantify the
information content of lump-type solutions in various scalar field
models, including two modified inverted $\phi^{4}$ models, the modified
$\phi^{3}$ model, as well as two additional families of lump models.
Our objective is to complement previous studies by providing an informational
perspective that distinguishes different solutions based on their
energy configurations. We explore how the DCC measure relates to energy
and its applicability in analyzing degenerate states. Our findings
indicate that DCC effectively correlates with the energy parameters
of the solutions, offering significant insights into their informational
properties. This study underscores the value of using informational
metrics like DCC to deepen our understanding of the structural and
dynamic characteristics of complex systems in theoretical physics.
\end{abstract}
\maketitle

\section{Introduction}

Configurational complexity (CC) refers to the degree of intricacy
in the spatial arrangement and interactions within a system. It encompasses
the diversity and distribution of states that the components of a
system can adopt, reflecting the system's organizational structure
and the relationships among its elements. In scientific terms, configurational
complexity is often quantified using metrics such as configurational
entropy (CE), which represents measure of information entropy that
pertains to the data compression and encoding processes by any source
\citep{Gleiser_PLB12,Gleiser_PRD12}. In other words, CE quantifies
the fraction of information encapsulated within the probability distributions
that characterize the underlying physical systems. High CC indicates
a large number of possible configurations and intricate interdependencies
among the components, leading to rich dynamical behavior and a high
level of unpredictability in the system's evolution.

In the realm of field theory, CC/CE and its generalizations has proven
instrumental in various analytical endeavors. Indeed, the continuum
differential variants, namely differential configurational entropy
(DCE) and differential configurational complexity (DCC), form the
foundational basis of configurational information measures. These
measures have emerged from the need to quantify the informational
content and structural complexity inherent in the configurations of
physical systems within the context of field theories \citep{Gleiser_PRD18}.
It has been leveraged to estimate critical phase transitions within
physical systems, delineate distinct regions within a system's phase
space, gauge the informational richness inherent in different solution
configurations, and elucidate disparities among configurations sharing
identical energy levels \citep{Gleiser_PRD12,Gleiser_PLB12,Gleiser_PLB13,Correa_PLB14,Gleiser_PRD15,Correa_AP15,Correa_EPJC15,Correa_EPJC16,Cruz_PLB17,Braga_PLB17,Bernardini_PLB17,Braga_PLB18,Braga_PLB18-2,Gleiser_PRD18,Lee_PLB19,Bazeia_JMMM19,Thakur_PLA20,Lee_PLB20,Thakur_PLA21,Koike_CSF22,Barreto_PRD22}.
This multifaceted approach underscores the utility of CC/CE in comprehensively
analyzing the intricate dynamics and structural nuances of complex
systems within the purview of theoretical physics.

Specifically, a measure of order, termed relative CE, was proposed
to quantify the emergence of coherent low-entropy configurations in
nonequilibrium field theory \citep{Gleiser_PRD12}. Also, in Ref.
\citep{Gleiser_PLB12}, this measure was applied to various nonlinear
scalar field models with spatially-localized energy solutions and
in Ref. \citep{Gleiser_PLB13} it was used to find the critical charge
for classically stable Q-balls and the Chandrasekhar limit. Moreover,
focusing on models exhibiting both double and single-kink solutions
treatable analytically via the Bogomol'ny-Prasad-Sommerfield bound,
in a parameter space where energy for distinct spatially-bound configurations
is degenerate, it was demonstrated in Ref. \citep{Correa_PLB14} that
CE effectively distinguishes between these energy-degenerate spatial
profiles. In Ref. \citep{Gleiser_PRD15} the authors demonstrate that
the critical stability regions of self-gravitating astrophysical objects
correlate closely with critical points of CE. The CE measure was also
used to study traveling solitons in Lorentz and CPT breaking scenarios
in Ref. \citep{Correa_AP15}, identifying the optimal parameter for
Lorentz symmetry breaking with symmetric energy density distribution.
Thick brane-world scenarios were examined in Ref. \citep{Correa_EPJC15}
and generalized theories of gravity were investigated using CE in
Ref. \citep{Correa_EPJC16}. Next, the CE framework was applied to
investigate the properties of degenerate Bloch branes in Ref. \citep{Cruz_PLB17}.
A technique was developed in Ref. \citep{Braga_PLB17} to define CE
for AdS-Schwarzschild black holes, showing its increase with temperature
and reliability as a measure of black hole stability. In Ref. \citep{Bernardini_PLB17}
the CE of glueball states was calculated in a holographic AdS/QCD
model, analyzing its dependence on glueball spin and mass to assess
state stability. In Ref. \citep{Braga_PLB18} the dissociation of
heavy vector mesons in a thermal medium was studied using CE in a
holographic AdS/QCD model, highlighting CE's role in assessing meson
stability against dissociation. A logarithmic measure of information,
using CE, was employed in Ref. \citep{Braga_PLB18-2} to quantitatively
study radially excited S-wave quarkonia states, revealing the relative
dominance and abundance of bottomonium and charmonium states, and
identifying the lower prevalence of higher S-wave resonances and heavier
quarkonia in nature. In Ref. \citep{Gleiser_PRD18} the authors used
differential CE to estimate oscillon lifetimes in scalar field theories
with symmetric and asymmetric double-well potentials and, in Ref.
\citet{Thakur_PLA20}, it was computed for solitons in tapered optical
waveguides and found that this measure saturates at a global minimum
value of its width. The CE of various tachyon kink solutions in tachyon
effective theory with Born-Infeld electromagnetic fields was investigated
in Ref. \citep{Lee_PLB19}. In Ref. \citep{Bazeia_JMMM19} the relationship
between the informational contents of spatially localized structures
and analytical solutions describing skyrmion-like structures in magnetic
materials was explored. The CE of a tachyonic braneworld with a bulk
cosmological constant was studied in Ref. \citep{Lee_PLB20}. Also,
the CC was computed for discrete soliton and rogue waves propagating
along an Ablowitz-Ladik-Hirota waveguide, modeled by a discrete nonlinear
Schrödinger equation in Ref. \citep{Thakur_PLA21}. Numerical investigation
of localized soliton-like solutions to a (2 + 1)-dimensional hydrodynamical
evolution equation was conducted in Ref. \citep{Koike_CSF22}. In
Ref. \citep{Barreto_PRD22} the gravitational instability of kinks
was analyzed using differential CE for both globally perturbed static
kinks and far-from-equilibrium kink solutions.

In the present paper, we utilize DCC to estimate the amount of information
stored in families of lump-type solutions, as studied in Refs. \citep{Avelar_EPJC08,Avelar_PLA09,Marques_EPL19},
with the aim of complementing those studies by providing an informational
perspective on distinguishing different solutions. These solutions
are indeed associated with certain parameters (details provided below)
that alter the energy configuration of the solutions. In this context,
we seek to understand how the DCC measure relates to energy and its
applicability in the case of degenerate states. As one knows, lumps
in scalar field theories have a direct connection with bright solitons,
which are important localized solutions that appear in nonlinear optics
\citep{Kivshar_03} and in Bose-Einstein condensates \citep{Pitaevskii_03}.
This possibility was discussed before in Refs. \citep{Avelar_EPJC08,Avelar_PLA09},
and in Refs. \citep{Belmonte-Beitia_PRL08,Avelar_PRE09,Cardoso_PRE13}
one finds a direct connection between bright solitons and lump-like
configurations. In this sense, the present investigation motivates
the study of the DCC measure for bright solitons in optical fibers
and in Bose-Einstein condensates.

The remainder of the paper is organized as follows. In the next section,
we present the methodology, including a review of the formalism for
a real scalar field and the structure of the DCC measure. The lump-type
models considered in this work are revisited in Section \ref{LM},
where we also present our results regarding their respective DCCs.
We conclude the present investigation in Section \ref{Conclusion},
where we comment on the main results and suggest new directions of
future investigations.

\section{Methodology}

\label{M}

\subsection{Scalar classical fields formalism}

From here we will review the formalism of field theory to get solutions
of localized structures like kinks and lumps, with the last one being
the target of this work. For this we consider the action 
\begin{equation}
S=\int\mathcal{L}\,d^{\mu}x,
\end{equation}
where the Lagrangian density of the model is given by 
\begin{equation}
\mathcal{L}=\frac{1}{2}\partial^{\mu}\phi\partial_{\mu}\phi-V(\phi),
\end{equation}
where $\phi$ describes a real scalar field, $\mu=0,1$ for $(1+1)$
space-time dimensions, and $V(\phi)$ is the potential. To get the
equation of motion we have to minimize the action $(\delta S=0)$,
which leads to 
\begin{equation}
\partial^{\mu}\partial_{\mu}\phi+V_{\phi}=0.\label{eom}
\end{equation}

In this work, we investigate field configurations using natural units,
with the metric such that $x^{\mu}=(x^{0}=t,x^{1}=x)$ and $x_{\mu}=(x_{0}=t,x_{1}=-x)$.
We consider the field, space, time and all the parameters included
in the system being rescaled to describe dimensionless quantities.
Since we are searching for static configurations for lump-like structures,
the above Eq. (\ref{eom}) takes the simpler form 
\begin{equation}
\phi{''}=V_{\phi},\label{eoms}
\end{equation}
with $\phi{''}=d^{2}\phi/dx^{2}$ and $V_{\phi}=dV/d\phi$. To find
lump-like solutions of Eq. (\ref{eoms}) we need the solutions satisfy
the boundaries conditions $\phi(x\to-\infty)=\phi(x\to\infty)=v$,
where $v$ is a local minimum of the potential. We can associate a
topological current $j^{\mu}=\epsilon^{\mu\nu}\partial_{\nu}\phi$
to this model, but for lumps the topological charge is zero.

The energy density $\rho$ can be obtained from the standard energy-momentum
tensor 
\begin{equation}
T^{\mu\nu}=\frac{\partial{\mathcal{L}}}{\partial{(\partial_{\mu}\phi)}}\partial^{\nu}\phi-\eta^{\mu\nu}\mathcal{L},
\end{equation}
where for $\mu=\nu=0$ one gets $\rho=T^{00}$. Considering that we
are working with static solution, we have 
\begin{equation}
\rho=\frac{1}{2}(\phi{'})^{2}+V.
\end{equation}

\begin{figure}[h]
\centering \includegraphics[width=0.8\columnwidth]{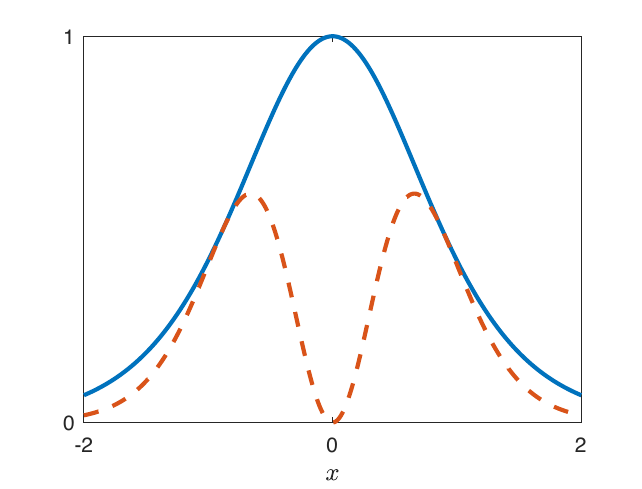}
\caption{The lump-like solution (blue solid line) and its energy density (orange
dashed line) for the $\phi^{3}$ model.}
\label{fig1} 
\end{figure}

As an example we review the $\phi^{3}$ model where we can obtain
standard lump-like solution. In this model the potential is given
by 
\begin{equation}
V(\phi)=2\phi^{2}(1-\phi),\label{phi3}
\end{equation}
which contains a minimum at $\phi=0$ (local minimum, with $V_{\phi\phi}({\phi=0})=4$),
and a zero at $\phi=1$. Thus, by using \eqref{phi3} into \eqref{eoms},
one gets 
\begin{equation}
\frac{d^{2}\phi}{dx^{2}}=2\phi(2-3\phi),
\end{equation}
which has the solution 
\begin{equation}
\phi(x)=\mathrm{sech}(x)^{2}\label{solphi3}
\end{equation}
and the corresponding energy density is given by 
\begin{equation}
\rho(x)=4\tanh(x)^{2}\mathrm{sech}(x)^{4}.\label{enphi3}
\end{equation}
Also, by integrating \eqref{enphi3} one gets the total energy $E={16}/{15}$.
The plot of the solution \eqref{solphi3} and energy density \eqref{enphi3}
are presented in Fig. \ref{fig1}. Note that in the peak of the lump
($x=0$) its energy density corresponds to $\rho=0$. For the maximum
of the density energy, we have $\rho=16/27\simeq0.592$ at $x\simeq\pm0.658$.

\subsection{Differential Configurational Complexity}

Following Ref. \citep{Gleiser_PLB12} and in view of that we are leading
with continuum systems, we will make a brief overview of the necessary
procedure to obtain the DCC, that is the continuum form of the CC.
First, as we are leading with lumps that has spatially-localized energy,
we have to consider functions that are bounded square-integrable $f(x)\in L^{2}(\mathbf{R})$.
The Plancherel's theorem states that 
\begin{equation}
\int_{-\infty}^{\infty}|f(x)|^{2}\,dx=\int_{-\infty}^{\infty}|F(k)|^{2}\,dk
\end{equation}
where $F(k)$ is the Fourrier transform of the $f(x)$. We can define
the DCC as 
\begin{equation}
\mathcal{C}_{C}[f]=-\int\tilde{f}(k)\ln[\tilde{f}(k)]\,d^{D}x,\label{DCC}
\end{equation}
where $D$ is related to the spatial dimensions, $\tilde{f}(k)=\frac{f(k)}{f_{max}(k)}$
and 
\begin{equation}
f(k)=\frac{|F(k)|^{2}}{\int|F(k)|^{2}\,d^{D}x}
\end{equation}
is the modal fraction. The term $f_{max}(k)$ will provide the maximal
modal fraction leading to $\tilde{f}(k)\leq1$ and ensuring that $\mathcal{C}_{C}[f]$
be positive-definite.

As demonstration we can obtain the DCC for the $\phi^{3}$ model discussed
above. With the energy density \eqref{enphi3} and using the expression
\eqref{DCC}, one calculate (numerically) the DCC that is given by
$\mathcal{C}_{C}(\rho)=2.3103$. For a better understanding of this
result we propose trial functions that approach and behave conform
to the boundaries conditions of the solution \eqref{solphi3}. First
we take $g(x)=\frac{1}{\sqrt{1+\lambda x^{2}}}$ where the total energy
$E$ is minimized for $\lambda\simeq11.628167$. Thus $E[g]=1.339107$
and the new DCC is $\mathcal{C}_{C}(g)=3.8064$. Let take another
trial function like $h(x)=\mathrm{sech}(\lambda x)$ and again, making
the same procedure, we find that now $\lambda=1.604749$ with $E[h]=1.069833$
and $\mathcal{C}_{C}(h)=2.4026$. We can notice that $\mathcal{C}_{C}(g)>\mathcal{C}_{C}(h)>\mathcal{C}_{C}(\rho)$
as expected due to higher energy (the more localized energy density)
higher the configurational complexity. It is worth mentioning that
we cannot guarantee that this result will always behave like this
for all trial functions that approach the solution of equation of
motion. Therefore, the DCC computes a measuring of complexity where
for high values we have more forms of configurations and the system
becomes more complex to describe in relation to the solution of the
equation of motion.

\section{Lump Models and its DCC}

\label{LM}

In this section we make a study of the DCC of some lump models developed
in Refs. \citep{Avelar_EPJC08,Avelar_PLA09,Marques_EPL19}. As we
will show below the DCC doesn't only provide a informational content
of dynamic of the localized structures but it can be used as a tool
to distinguish configurations with degenerate energy.

\subsection{Modified inverted $\phi^{4}$ model}

\begin{figure}[tb]
\centering \includegraphics[width=0.8\columnwidth]{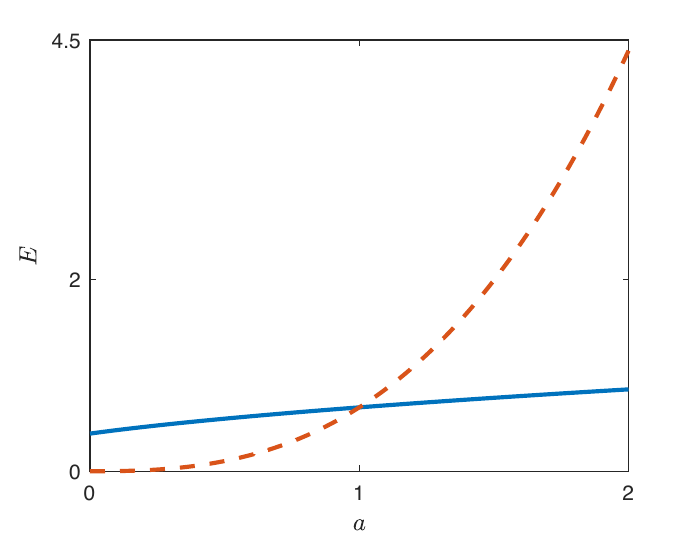}
\includegraphics[width=0.8\columnwidth]{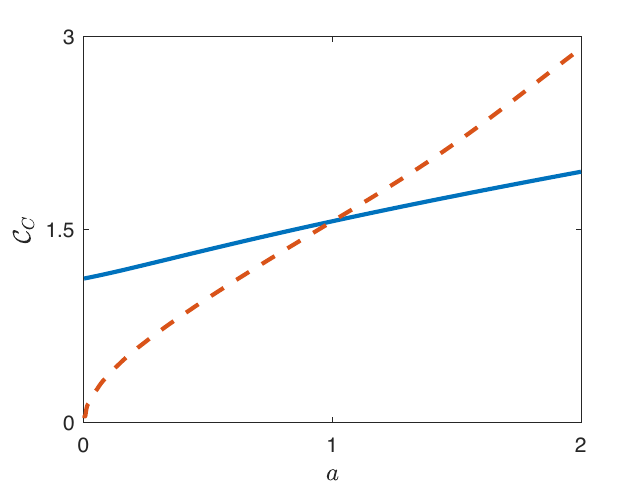} \caption{Energy (top) and DCC (bottom) of the modified inverted $\phi^{4}$
model, for the two sectors: blue (solid) line (sector $1$) and orange
(dashed) line (sector $a$). The energy degenerates for $a=1$, as
expected.}
\label{Ephi4invertido} 
\end{figure}

Let us start with the modified inverted $\phi^{4}$ model described
in \citep{Avelar_EPJC08}. This model has the potential

\begin{equation}
V(\phi)=\frac{1}{2}\phi^{2}(1+\phi)(a-\phi),\label{p41}
\end{equation}
where $a$ is a positive parameter and the zeros of this potential
are $\phi=0$ (local minimum, with $V_{\phi\phi}(\phi=0)=a$, and
$\phi=-1$ and $\phi=a$. The lump-like solutions of this model are
given by

\begin{align}
\phi_{1}(x) & =\frac{2a}{1-a-(1+a)\cosh(\sqrt{a}x)}\\
\phi_{a}(x) & =\frac{2a}{1-a+(1+a)\cosh(\sqrt{a}x)}
\end{align}
which are related to two sectors of the potential, sector 1 and sector
a, with the solutions $\phi_{1}$ and $\phi_{a}$. The corresponding
energy densities are

\begin{align}
\rho_{1}(x) & =\frac{4a^{3}(1+a)^{2}\sinh(\sqrt{a}x)^{2}}{\PC{1-a-(1+a)\cosh(\sqrt{a}x)}^{4}},\\
\rho_{a}(x) & =\frac{4a^{3}(1+a)^{2}\sinh(\sqrt{a}x)^{2}}{\PC{1-a+(1+a)\cosh(\sqrt{a}x)}^{4}}.
\end{align}
After integrating them in all space, we get the total energies

\begin{align}
E_{1}(a) & =\frac{\pi}{8}(1-a)(1+a)^{2}+E_{a},\\
E_{a}(a) & =-\frac{1}{4}(1-a)(1\!+a)^{2}\arctan(\sqrt{a})\!+\!\frac{1}{12}\sqrt{a}(3\!+\!2a\!+\!3a^{2}).
\end{align}
One notices that in the limit $a\to1$, the above solutions become
$\phi=\pm\mathrm{sech}(x)$ and the two energies converges to the
single value $E={2}/{3}$. By using the energy densities, one can
obtain the corresponding DCC. The results are displayed in Fig. \ref{Ephi4invertido},
showing that the only point in which the energy degenerates is $a=1$
and at the same point, the solutions has equal DCC. Therefore, we
cannot use the DCC to discriminate the degenerate energy for this
case but we notice that for the region $0<a<1$ we have $E_{\phi_{1}}>E_{\phi_{a}}$
with $\mathcal{C}_{C_{\phi_{1}}}>\mathcal{C}_{C_{\phi_{a}}}$ and
for $a>1$ we have $E_{\phi_{1}}<E_{\phi_{a}}$ with $\mathcal{C}_{C_{\phi_{1}}}<\mathcal{C}_{C_{\phi_{a}}}$,
an expected result as discussed before: the higher the energy is,
the higher the DCC also is. In the next example we see that this behavior
does not always happen.

\subsection{Modified $\phi^{4}$ model}

Now we consider the modified $\phi^{4}$ model introduced in \citep{Avelar_EPJC08}.
In this case the potential is written as 
\begin{equation}
\begin{aligned}V(\phi) & =2\phi^{2}(\phi-b)\left(\phi-\frac{b}{c}\right)\\
 & =2\phi^{2}(\phi-\tanh(a))(\phi-\coth(a)),
\end{aligned}
\label{p42}
\end{equation}
where $b>0$, $c\geq1$ and we consider $b=\tanh(a)$ and $c=\tanh(a)^{2}$.
This potential has 3 zeros for $\phi=0$ (local minimum) with $V_{\phi\phi}=\frac{4b^{2}}{c}$,
and $\phi=b$ and $\phi=\frac{b}{c}$. The lump-like solution is given
by 
\begin{equation}
\phi(x)=\frac{1}{2}[\tanh(a+x)-\tanh(a-x)],
\end{equation}
and the energy density with total energy is given, respectively, by
\begin{align}
\rho(x) & =\frac{1}{4}[\tanh(a+x)^{2}-\tanh(a-x)^{2}]^{2},\\
E(a) & =\frac{2}{3}-2\text{cosech}(2a)^{2}(2a\coth(2a)-1).
\end{align}

\begin{figure}[tb]
\centering \includegraphics[width=0.8\columnwidth]{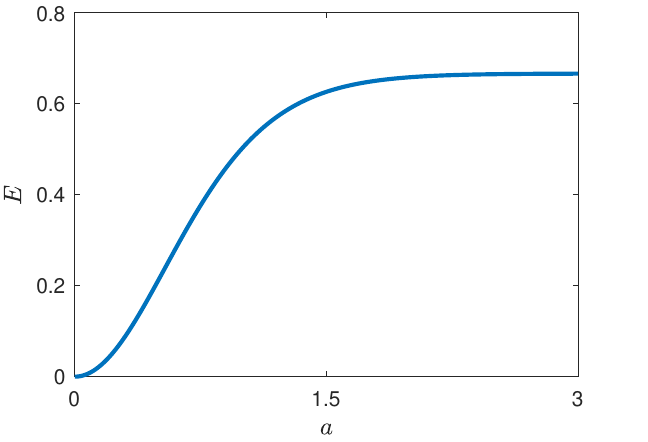}
\includegraphics[width=0.8\columnwidth]{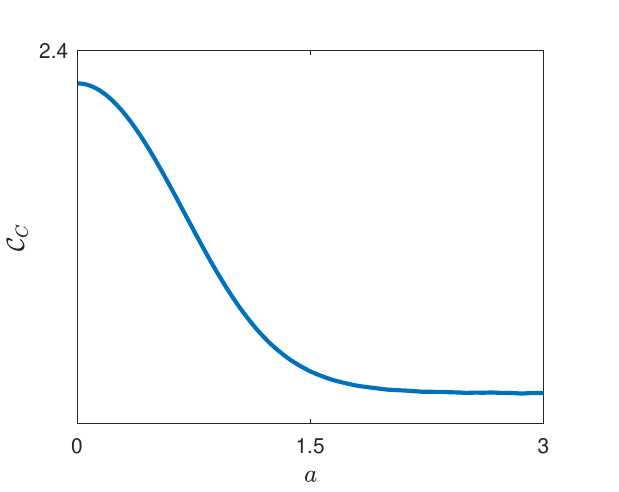} \caption{Energy (top) and DCC (bottom) of the modified $\phi^{4}$ model. As
the parameter $a$ increases the energy (DCC) increases (decreases),
until a certain point. After that point they become constant.}
\label{E_phi4imodificado} 
\end{figure}

We display the total energy $E$ and the DCC for this model in Fig.
\ref{E_phi4imodificado}. Note that, unlike the modified inverted
$\phi^{4}$ model, in this case the energy increases and the DCC decreases
as one increases $a$. When $a$ increases to higher and higher values
both the energy and DCC tend to saturate at some constant values.
As one sees, the complexity of the solution decreases as the parameter
$a$ increases.

\subsection{The modified $\phi^{3}$ model}

\label{p3 model}

We consider another model, previously investigated in Ref. \citep{Avelar_EPJC08},
the so-called modified $\phi^{3}$ model. The potential is given by
\begin{equation}
V(\phi)=2p^{2}\phi^{2-\frac{2}{p}}(1-a-\phi^{\frac{1}{p}})(a+\phi^{\frac{1}{p}})^{2}
\end{equation}
where $a$ and $p$ are positive parameters with $a\in[0,1]$ and
$p$ is odd integer. These parameters control the features of the
lump. If we make $a\to0$ and $p\to1$ we get back the $\phi^{3}$
model already discussed. The zeros of this potential are $\phi=0$,
$\phi=-a^{p}$ (local minimum) and $\phi=(1-a)^{p}$. The lump-like
solution of this model is 
\begin{equation}
\phi(x)=(\mathrm{sech}(x)^{2}-a)^{p},
\end{equation}
with the corresponding energy density 
\begin{equation}
\rho(x)=4p^{2}\mathrm{sech}(x)^{4}\tanh(x)^{2}[\mathrm{sech}(x)^{2}-a]^{2p-2}.\label{edmp3}
\end{equation}

To find the expression for the total energy of \eqref{edmp3} we first
consider the case for $p=3$. Thus, we have 
\begin{equation}
E(a)=\frac{2048}{1001}-\frac{4096}{385}a+\frac{768}{35}a^{2}-\frac{768}{35}a^{3}+\frac{48}{5}a^{4}.\label{enPhimod1}
\end{equation}

\begin{figure}[tb]
\centering \includegraphics[width=0.8\columnwidth]{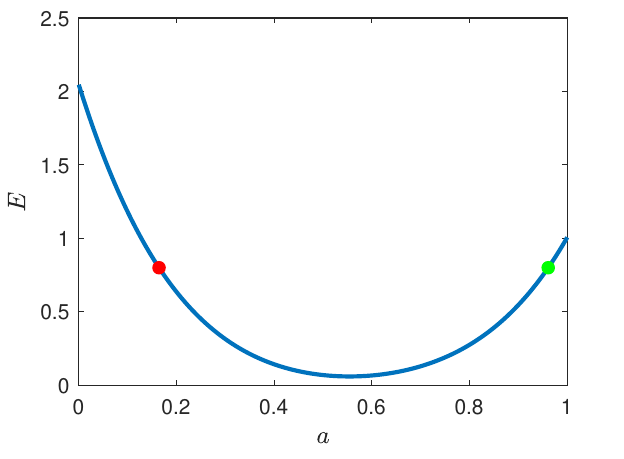}
\includegraphics[width=0.8\columnwidth]{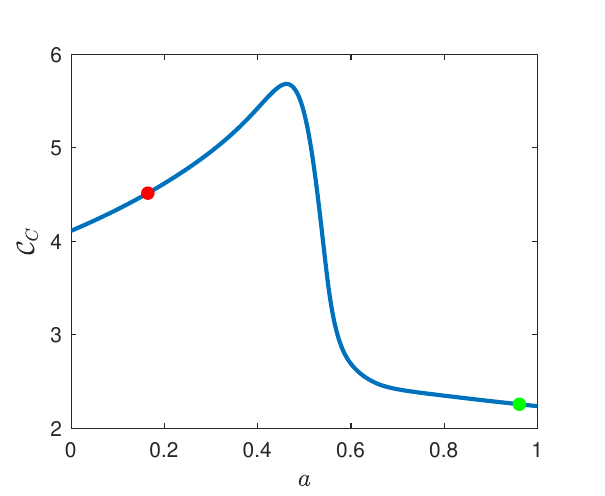}
\caption{Energy (top) and DCC (bottom) for the modified $\phi^{3}$ model,
with $p=3$. The red and green dots mark the same energy value $E(a)=0.8$
in the top panel, and distinct values for the DCC in the bottom panel,
respectively, $\mathcal{C}_{C}=4.5149$ and $\mathcal{C}_{C}=2.2555$. }
\label{EnPhi3mod1} 
\end{figure}

We plot the energy (\ref{enPhimod1}) and the DCC for this model,
with $p=3$. We can see from Fig. \ref{EnPhi3mod1} that the energy
has an infinity of values of degenerate energy. We choose to analyze
the degenerate energy for $E(a)=0.8$, where the corresponding values
for $a$ are $a_{1}\simeq0.1642967598$ (red dot) and $a_{2}\simeq0.9610022128$
(green dot). In the figure, these points corresponding to $\mathcal{C}_{C}(a_{1})=4.5149$
(red dot) and $\mathcal{C}_{C}(a_{2})=2.2555$ (green dot). Thus,
with $\mathcal{C}_{C}(a_{1})>\mathcal{C}_{C}(a_{2})$, we can say
that the point $a_{1}$ has a high degree of complexity than the point
$a_{2}$ for the same value of energy. Therefore, the DCC may act
to discriminate distinct states of same energy.

\begin{figure}[tb]
\centering \includegraphics[width=0.85\columnwidth]{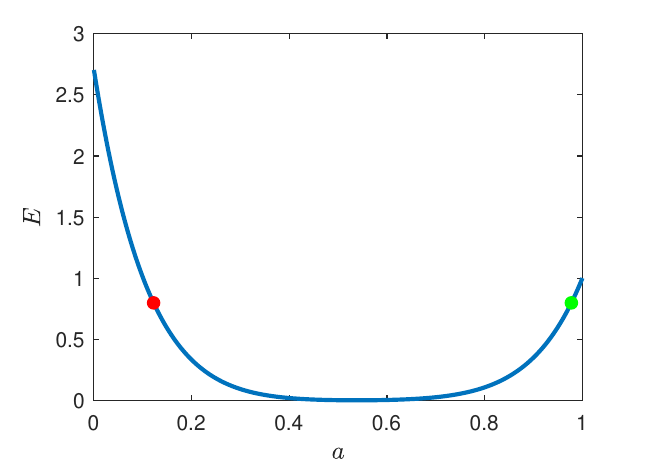}
\includegraphics[width=0.8\columnwidth]{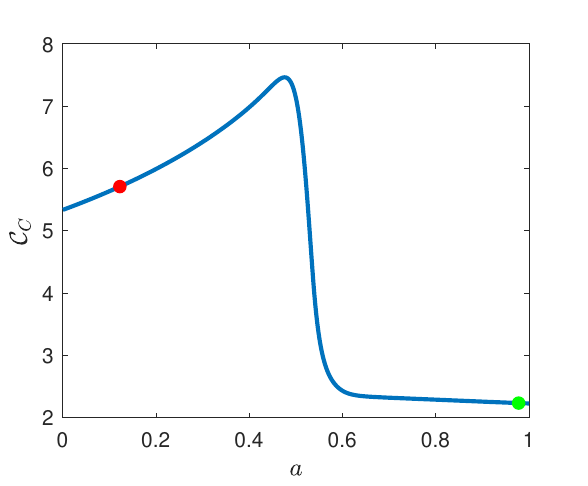}

\caption{Energy (top) and DCC (bottom) for the modified $\phi^{3}$ model,
with $p=5$. The red and green dots mark the same energy value $E(a)=0.8$
in the top panel, and distinct values for the DCC in the bottom panel,
respectively, $\mathcal{C}_{C}=5.7103$ and $\mathcal{C}_{C}=2.2349$.}
\label{EnPhi3mod} 
\end{figure}

Similar procedure can be done for the case with $p=5$, but now the
energy is given by 
\begin{equation}
\begin{aligned}E(a) & =\frac{2621440}{969969}-\frac{10485760}{415701}a+\frac{2293760}{21879}a^{2}-\frac{327680}{1287}a^{3}\\
 & +\frac{512000}{1287}a^{4}-\frac{40960}{99}a^{5}+\frac{2560}{9}a^{6}-\frac{2560}{21}a^{7}+\frac{80}{3}a^{8},
\end{aligned}
\end{equation}
where we choose the energy value $E(a)=0.8$. For this value we have
two points, $a_{1}=0.1221365572$ (red point) and $a_{2}=0.9774293631$
(green point) and the DCC are $\mathcal{C}_{C}(a_{1})=5.7103$ and
$\mathcal{C}_{C}(a_{2})=2.2349$. We display the curves of $E$ and
$\mathcal{C}_{C}$ in Fig. \ref{EnPhi3mod}. We can use again, for
this case ($p=5$), the DCC to distinguish the degenerate energy value
$E=0.8$, where we can see that $\mathcal{C}_{C}(a_{1})>\mathcal{C}_{C}(a_{2})$.
Observing in the DCC curve, the points in both cases ($p=3$ and $p=5$),
we notice that the green point has less complexity than the red point.
But comparing the red point of both, the DCC in the case for $p=5$
is higher. For the green point, in both cases, we can see the opposite,
the DCC for the case $p=5$ is slightly below. Anyway, we can use
the DCC to choose the solution that engenders less or more complexity,
even when they have the same energy value.

In the following two subsections, we further examine two additional
lump models. The first, as introduced in Ref. \citep{Avelar_PLA09},
involves a potential characterized by an inverted $\phi^{n+2}$ term.
The second, detailed in Ref. \citep{Marques_EPL19}, is a two-parameter
model ($n$ and $m$) that yields solutions featuring polynomial tails.

\subsection{Lump in inverted $\phi^{n+2}$ model}

Now we consider the following potential \citep{Avelar_PLA09} 
\begin{align}
V(\phi)=\frac{2}{n^{2}}\phi^{2}(1-\phi^{n}),
\end{align}
where $n$ is a positive integer and the zeros of this potential,
for odd values, are $\phi=0$ and $\phi=1$. For even values, they
are $\phi=0$ and $\phi=\pm1$.

The solution of this model is given by 
\begin{equation}
\phi(x)=\mathrm{sech}(x)^{\frac{2}{n}}
\end{equation}
and the corresponding energy density is 
\begin{equation}
\rho(x)=\frac{4}{n^{2}}\mathrm{sech}(x)^{\frac{4}{n}}\tanh(x)^{2}.
\end{equation}
Integrating it, we get the total energy as 
\begin{equation}
E(n)=\frac{\sqrt{\pi}}{2n}\frac{\Gamma(\frac{n+2}{n})}{\Gamma(\frac{3n+4}{n})}.
\end{equation}

\begin{figure}[tb]
\centering \includegraphics[width=0.8\columnwidth]{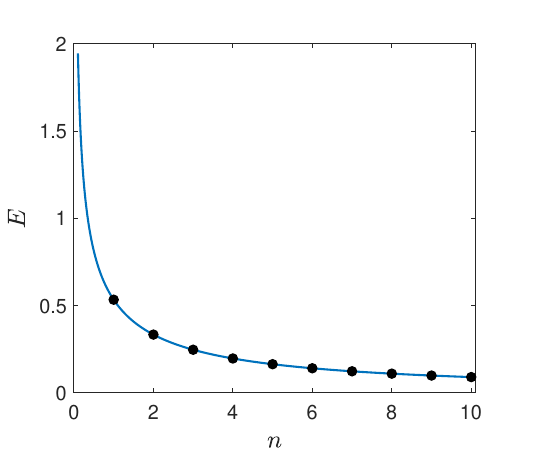}
\includegraphics[width=0.8\columnwidth]{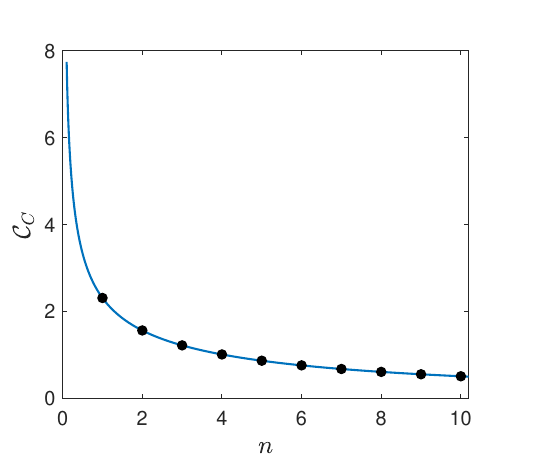} \caption{Energy (top) and DCC (bottom) of the inverted $\phi^{n+2}$ model,
with the black points representing $n=1,2,...,10$. The blue thin
line serves to indicate the asymptotic behavior the two quantities.}
\label{primeira_familia_E} 
\end{figure}

Next, we plot the total energy $E$ and the DCC of this model in Fig.
\ref{primeira_familia_E}. Observe that there are no degenerate values
of the energy and the behavior of the DCC is the expected by what
was discussed before when the energy increases (decreases) the DCC
increases (decreases). Indeed, we observe here that as the energy
$E$ approaches zero, the DCC correspondingly tends to zero, i.e.,
$\mathcal{C}_{C}\to0$.

\subsection{Lump with power law tail}

The other model is related with lumps with power law tail introduced
in Ref. \citep{Marques_EPL19}. In this case, the two-parameter potential
is expressed as follows: 
\begin{equation}
V(\phi)=2\frac{n^{2}}{m^{2}}\phi^{2+\frac{m}{n}}(1-\phi^{m})^{2-\frac{1}{n}},
\end{equation}
where $m$ is a positive even number and $n$ is a positive odd parameter.
The solution is given by 
\begin{equation}
\phi(x)=\frac{1}{(1+x^{2n})^{\frac{1}{m}}},
\end{equation}
and the corresponding energy density is 
\begin{equation}
\rho(x)=4\frac{n^{2}}{m^{2}}\frac{x^{4n-2}}{(1+x^{2n})^{2+\frac{2}{m}}}.
\end{equation}

\begin{figure}[tb]
\centering \includegraphics[width=0.78\columnwidth]{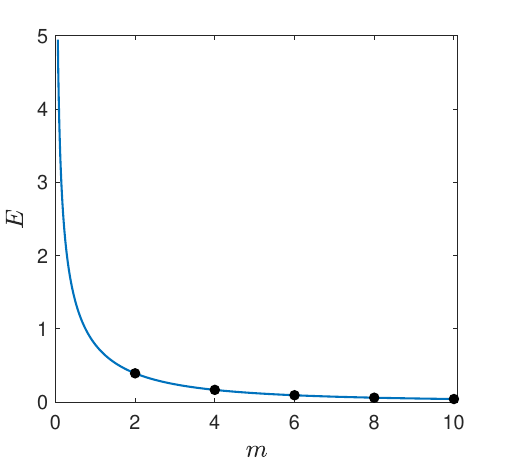}
\includegraphics[width=0.8\columnwidth]{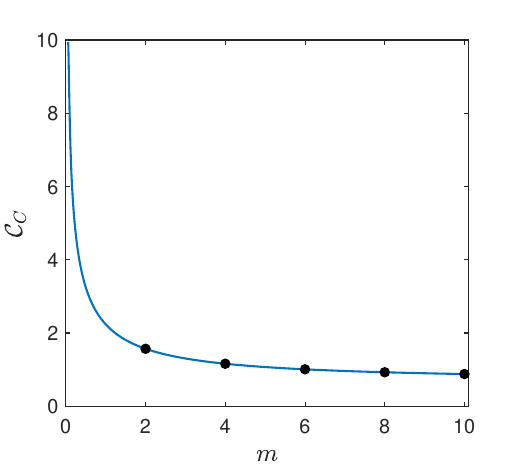} \caption{Energy (top) and DCC (bottom) for the power law model with the parameter
$n=1$ and $m=2,4,6,8,$ and $10$. We see that the energy vanishes
as $m$ increases to higher and higher values. The blue thin line
serves to indicate the asymptotic behavior of the two quantities.}
\label{long_tail_E} 
\end{figure}

A general expression for the total energy is not analytical and we
just consider the case for $n=1$ and $m$ even integer. Thus, we
get 
\begin{equation}
E(m)=\frac{2\sqrt{\pi}}{m^{2}}\frac{\Gamma(\frac{m+4}{2m})}{\Gamma(\frac{2(m+1)}{m})}.
\end{equation}
We show the total energy and DCC of this model in Fig. \ref{long_tail_E}.
By setting $n=1$, we observe that $E\to0$ as $m$ increases to large
values, and the DCC converges to the value $\mathcal{C}_{C}=0.6733$.
We also conducted an analysis of the DCC behavior for additional values
of $n$ and observed that it exhibits a pattern (not shown here) with
limiting values of DCC as $m\to\infty$ differing from those obtained
when $n=1$. Specifically, for instance, we obtained approximately
$\mathcal{C}_{C}=0.9026$ for $n=2$, $1.2642$ for $n=4$, $1.5067$
for $n=6$, and $1.7734$ for $n=10$.

\section{Conclusion}

\label{Conclusion}

In this work, we have demonstrated the utility of differential configurational
complexity in estimating the information content of lump-type solutions
across several scalar field models, specifically the two modified
inverted $\phi^{4}$ models, the modified $\phi^{3}$ model, the inverted
$\phi^{n+2}$ model, and the two parameter model presenting lumps
with power law tail. Our analysis reveals that DCC is a robust measure
for distinguishing between different energy configurations and provides
significant insights into the informational aspects of these solutions.
Notably, it has been observed that the DCC measure may correlate with
the energy parameters of the solutions, underscoring its potential
utility in the analysis and characterization of degenerate states.
However, it is important to note that the DCC measure is model-dependent,
as its behavior does not always exhibit a pattern directly aligned
with the energy curve.

In particular, an interesting result is directly related to the modified
$\phi^{3}$ model studied in Sec. \ref{p3 model}, which shows that
the energy degeneracy of the solution is not observed in the corresponding
DCC curve. In this sense, the DCC may prove useful in distinguishing
between solutions with the same energy, but with distinct DCC values.
This result motivates us to examine the DCC measure for bright solitons
in optical fibers and in Bose-Einstein condensates, which appear from
the nonlinear Schrödinger equation \citep{Kivshar_03,Pitaevskii_03}.
Furthermore, we are now embarking on new studies, to understand the
behavior and correlation of the DCC measure with the energy of lump-like
solutions in systems described by two or more scalar fields, with
the additional fields serving to describe new degrees of freedom of
the corresponding models. A possibility of practical interest concerns
the presence of twistons in polyethylene \citep{Mansfield_JPS78}.
In the work \citep{Bazeia_CPL99}, the study of twistons considered
two real scalar fields, the second field used to describe the rotational
degree of freedom associated to the topological configurations. We
think the study of the DCC for twistons in polyethylene may perhaps
bring new information to this kind of topological configuration. Similar
investigation can be carried out in the case of localized structures
for proton conductivity in langmuir films \citep{Bazeia_CPL01}. We
hope to present new results in an upcoming paper.

\subsection*{Acknowledgements}

The authors acknowledge the financial support of the Brazilian agencies
CNPq (\#303469/2019-6, \#402830/2023-7, \#306105/2022-5, \#407469/2021-4,
\#312173/2022-9), Sisphoton Laboratory-MCTI (\#440225/2021-3), FAPEG
(\#202110267000415) and CAPES. This work was also performed as part
of the Brazilian National Institute of Science and Technology (INCT)
for Quantum Information (\#465469/2014-0). Thanks are also due to
Paraiba State Research Foundation, FAPESQ-PB (\#0015/2019).

\bibliographystyle{apsrev4-2}
\bibliography{bibliography}

\end{document}